# Accurate Sky Continuum Subtraction with Fibre-fed Spectrographs

Yanbin Yang[1]
Myriam Rodrigues[2]
Mathieu Puech[1]
Hector Flores[1]
Frederic Royer[1]
Karen Disseau[1]
Thiago Gonçalves[3]
François Hammer[1]
Michele Cirasuolo[4]
Chris Evans[5]
Gianluca Li Causi[6]
Roberto Maiolino[7]
Claudio Melo[2]

[1] GEPI, Observatoire de Paris, CNRS, Université Paris Diderot, Meudon, France
[2] ESO
[3] Observatorio do Valongo, Brazil
[4] SUPA, Institute for Astronomy, University of Edinburgh, Royal Observatory of Edinburgh, UK
[5] UK Astronomy Technology Centre, Royal Observatory of Edinburgh, UK
[6] INAF, Osservatorio Astronomico di Roma, Italy
[7] Cavendish Laboratory, University of Cambridge, UK

Fibre-fed spectrographs now have throughputs equivalent to slit spectrographs. However, the sky subtraction accuracy that can be reached has often been pinpointed as one of the major issues associated with the use of fibres. Using technical time observations with FLAMES–GIRAFFE, two observing techniques, namely dual staring and cross beam-switching, were tested and the resulting sky subtraction accuracy reached in both cases was quantified. Results indicate that an accuracy of 0.6 % on sky subtraction can be reached, provided that the cross beam-switching mode is used. This is very encouraging with regard to the detection of very faint sources with future fibre-fed spectrographs, such as VLT/MOONS or E-ELT/MOSAIC.

## Why fibres could be an issue when observing faint targets

One of the key science drivers for the future instrumentation of the Very Large Telescope (VLT) and the European Extremely Large Telescope (E-ELT) is to study faint galaxies in the early Universe at very high redshifts (e.g., Navarro et al., 2010; Cirasuolo et al., 2011). The detection and spectroscopic follow-up of these sources will require an accurate and precise sky subtraction process. For instance, in its deepest observations, VLT/MOONS (Cirasuolo et al., 2011) will study sources as faint as $H_{AB}$ = 25 mag in their continuum and emission lines in ~ 16 hours of integration, while E-ELT/MOSAIC (which is a concept design for a multi-object spectrograph [MOS] on the E-ELT; Evans et al., in prep.) will push this limit up to $J/H_{AB}$ ~ 30 mag in emission, and up to $J/H_{AB}$ ~ 27 mag for continuum and absorption line features. These spectral features will be typically observed between bright OH sky lines. However, the near-infrared (NIR) sky continuum background is found to be $J/H_{AB}$ ~ 19–19.5 mag in dark sky conditions (Sullivan & Simcoe, 2012), i.e., hundreds to a thousand times brighter than the sources to be detected in the continuum. While the expected relatively bright emission lines with restframe equivalent widths larger than ~ 15 nm (e.g., Navarro et al., 2010) should easily emerge above this background, continuum and absorption line detections will clearly be hampered by possible systematic residual signal left by the sky subtraction process. For the future detection of such faint sources, it is therefore critical to check that sky subtraction techniques are accurate enough, i.e., that one can actually reach accuracies at a level of a few tenths of a percent at least.

In this respect, slit spectrographs have long been considered as much more accurate than fibre-fed spectrographs. This is mainly due to two different issues: (i) fibre-fed spectrographs, if not carefully designed, can suffer significant loss of light compared to slit spectrographs (resulting from, e.g., fibre cross-talk on the detector or focal ratio degradation [FRD] which results in a change of aperture at the output of the fibres); (ii) it is more difficult to achieve an accurate sky subtraction process with fibres than with slits. Recent developments in fibre technology and careful designs of fibre-fed spectrographs can control the issues of the first kind. For instance, good spacing of fibres on the detector can avoid significant cross-talk, while fibre-fed spectrographs can now potentially reach global throughputs similar to those of classical multi-slit spectrographs (e.g., Navarro et al., 2010).

However, issues related to the accuracy of sky subtraction remain, and this turns out to be particularly problematic when dealing with faint sources. Perhaps the two most important effects often associated with fibres, which further limit the accuracy of the sky subtraction compared to slits, are: (1) sky continuum variations between the position of the object and the position at which the sky is measured (due to the finite coverage of the fibres in the focal plane and the minimum practical distance of closest approach); and (2) variations in the fibre-to-fibre response (due, e.g., to point spread function [PSF] variations, fringing or FRD). We refer to, e.g., Sharp & Parkinson (2010) for a detailed description of the factors limiting the sky subtraction accuracy with fibres. After the E-ELT phase A instrumentation studies finished, we undertook to better characterise these two important potential caveats. We now report on some of the results obtained during the past two years.

## Characterising the sky continuum spatial variations

The signal from the sky is difficult to predict and subtract in the NIR, mainly because of its strong variability in space and time. Variations of about 15 % in amplitude are typical over spatial scales of a few degrees (e.g., Moreels et al., 2008). These variations are clearly dominated by the flux fluctuations of the bright and numerous OH emission lines, and, in second order, by the intensity variations of absorption bands produced by molecules of water and other components of the Earth's atmosphere. Between these telluric emission and absorption lines, the sky signal has a continuum level about ~ 19–19.5 $mag_{AB}$ in the $J/H$ bands, the origin of which is still poorly understood. It could be due to a pseudo-continuum associated with instrumental residuals, e.g., diffuse scattered light from the wings of bright emission lines within the spectrograph (Trinh et al., 2013), or to continuum radiation from constituents of the atmosphere. To our knowledge, the spatial and temporal variability of this sky continuum as a function of both space and time has never been characterised.

We first investigated these issues using archival VLT/FORS2 narrowband imaging and spectroscopy data. Over timescales of a few tens of minutes, Puech et al. (2012) and Yang et al. (2012) found that the sky continuum background exhibits spatial variations over scales from ~ 10 to ~ 150 arcseconds, with total amplitudes below 0.5% of the mean sky background. At scales of ~ 10 arcseconds or below (on which the sky is likely to be measured with future fibre-fed spectrographs), the amplitude of the variations is found to be ~ 0.3–0.7%. Note that this should still be considered as an upper limit to real sky continuum variations, since scattered light and noise variance can be difficult to mitigate in such low signal-to-noise data. Regardless, this is a very encouraging result, which suggests that sky background subtraction can, in principle, be achieved with an accuracy of a few tenths of a percent.

### Testing sky subtraction with cross beam switching observation

Reaching sub-percent accuracy on sky subtraction still requires dealing with the variations in fibre-to-fibre response. For this purpose we requested ESO technical time on FLAMES–GIRAFFE, which is the ESO optical workhouse multi-object fibre-fed instrument at VLT/UT2. Such tests reveal that accuracies of few tenths of a percent can indeed be reached, provided that a cross beam-switching observing sequence is used (see Figure 1). In the following, we describe the observations conducted and the results obtained.

We undertook technical observation with FLAMES–GIRAFFE on 8 March 2012, using the Medusa mode with clear conditions and a seeing ~ 0.9 arcseconds. The fibres were distributed over a 20 by 20 arcminute region in the zCOSMOS field (Lilly et al., 2007). Seventy fibres were distributed in pairs separated by 12 arcseconds and oriented along the north–south axis, as illustrated in Figure 1. Three of these pairs were "pure sky", meaning that no object was observed in any pair of fibres. Preliminary results from these three pairs were presented in Rodrigues et al. (2012).

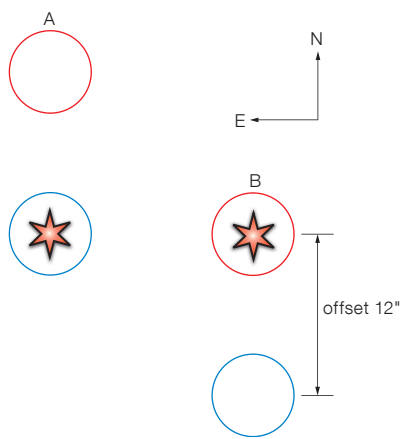

Figure 1. Illustration of cross beam-switching observations. The astrophysical object of interest is represented as a star. The fibres are positioned in pairs in the focal plane, with distances of 12 arcseconds between the two fibres of each pair along the north–south axis. Such a pair is represented by red and blue circles. The observing sequence consists of dithering the telescope from position A to position B by 12 arcseconds such that the object (and the sky) is always observed within one of the two fibres of a given pair.

The LR8 GIRAFFE setup, which covers 820–940 nm with a spectral resolution of $R = 6500$, was chosen to obtain spectra at NIR wavelengths. The target field was observed at low airmass (< 1.2), i.e., when it was relatively close to the meridian with an hour angle of less than ~ 30 minutes. During the observations, the Moon was located ~ 28 degrees away from the target field, contributing about 50% to the sky continuum background flux. The total continuum background of the observations reaches $m_{AB}$ ~ 19.7 mag arcsecond$^{-2}$, which is very similar to the $J$-band sky continuum brightness in dark conditions (Sullivan & Simcoe, 2012) and therefore particularly well-suited to our purposes.

The observations were carried out using a cross beam-switching configuration in which the telescope was offset by 12 arcseconds along the north–south axis three times in a row (see Figure 1), resulting in an A–B–A–B–A–B sequence. FLAMES does not have a template for cross beam-switching observations, so the telescope guiding had to be switched off during all the dithered B exposures. In principle, the pointing error during offsets is better than 0.2 arcseconds, which is much smaller than the diameter of the fibres (1.2 arcseconds). However, in order to prevent any risk of significant misalignments between the objects and the fibres during the dithered B exposures, an A–B–A–B-like sequence was preferred instead of the usual A–B–B–A-like sequence. This preserved the signal-to-noise ratio and only resulted in larger overheads. Each individual (A or B) exposure was ten minutes. The three consecutive A–B sequences obtained represent a total effective exposure time of one hour, immediately after which attached flat-field exposures were acquired.

### Data analysis and results

Basic reduction steps were performed using the ESO pipeline (bias correction, internal calibration lamp flat-fielding, wavelength calibration and extraction of 1D spectra). We focused on the sky continuum background since targets will be observed between the sky emission and strong telluric lines (e.g., Vacca et al., 2003; Davies et al., 2007). Seven spectral regions free of sky emission and absorption lines were defined (see Figure 2) to test the accuracy of different sky subtraction strategies. To increase the statistics, but limit the impact of the object spectrum, we limited the analysis to 15 pairs with object $I_{AB}$-band magnitudes fainter than 21. The mean magnitude of these 15 objects is found to be $I_{AB} = 21.88$ mag, which corresponds to a continuum flux that is ~ 7 times fainter than the contribution from the sky continuum. To compare the sky continuum subtraction accuracy reached using cross beam-switching observations, we also reduced the data in order to mimic a simpler staring mode observing strategy. Both observing and reduction methods are detailed below.

In staring mode observations, an object and the nearby sky (12 arcseconds away in the present case) are observed with a pair of fibres simultaneously. For each object in the sample, we derived two spectra by combining the three A exposures and the three B exposures, respectively. The resulting exposure time of each spectrum is 30 minutes. Here, there is no point in combining all A and B exposures together since, by definition, in staring mode the sky is sampled on only one side of the object, and the object and sky are observed with different fibres which would lead to large residuals. Thus, we simply subtracted the integrated B spectrum from the integrated A spectrum. Since the objects have fluxes that are significantly fainter than that of the sky continuum (see above), the resulting difference can be considered as a first-order estimate of the residuals from the sky subtraction process.

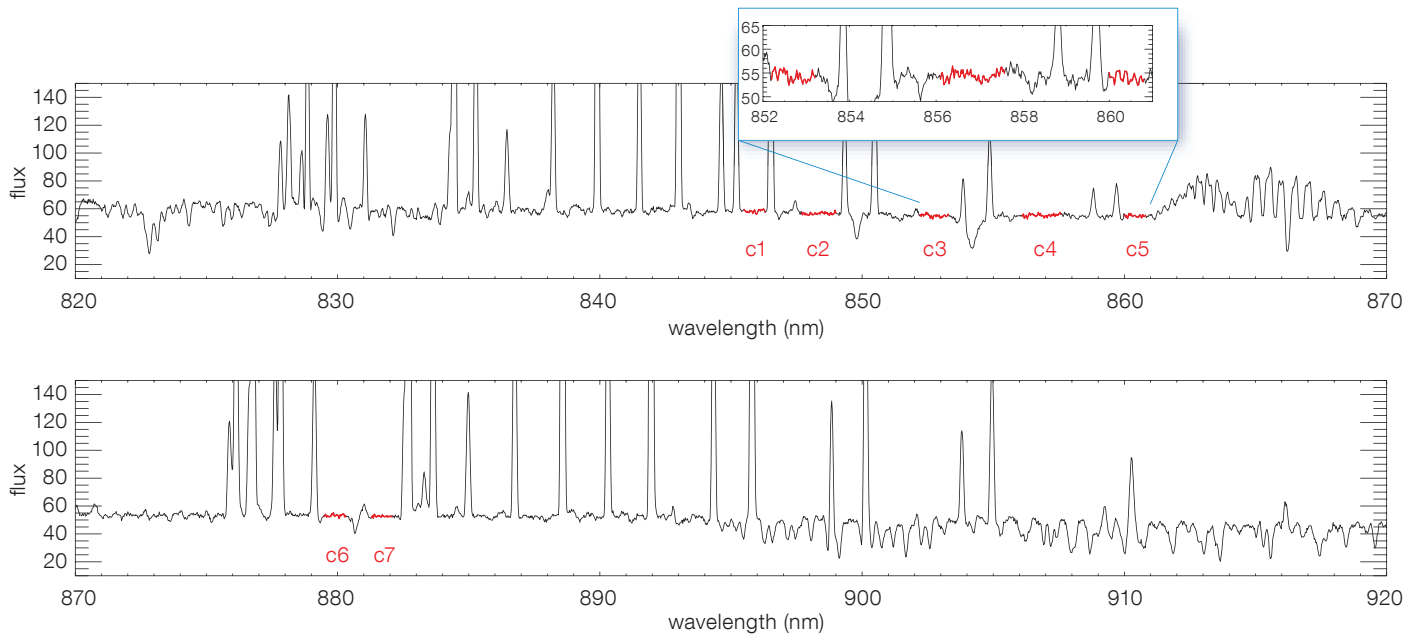

Figure 2. Sky spectrum from 820 to 920 nm obtained after combining 72 sky fibres. OH emission lines and atmospheric absorptions, i.e., telluric lines, can be seen almost everywhere. Only a few relatively clean continuum regimes are left, such as the red regions marked as $c_1$ to $c_7$, which are selected for further analysis.

In cross beam-switching mode, the telescope is dithered by 12 arcseconds along the north–south axis between the A and B positions. During the three consecutive A–B sequences, a given object is always observed by one of the fibres of the pairs alternately. Within a single A–B sequence, the object spectrum can be extracted twice (once in A and once in B) and subtracted one from another. In contrast to the staring mode, one can combine all six exposures to produce a spectrum with a one-hour integration. In cross beam-switching mode, object and sky are indeed observed by the same fibre but at different times, and the sky is sampled at both sides around the objects. Finally, if one combines all the exposures of all the objects together, one can actually simulate the result of a 15-hour on-sky spectrum. Combining all the measurements in the seven distinct spectral windows can mimic a simulated integration time of up to 105 hours. This is sufficiently long so as to sample deep integration times for future VLT or E-ELT observations (i.e., of a few tens of hours). In the absence of any residual systematic effect, one would expect that the resulting signal-to-noise ratio of the combined spectrum is reduced by a factor 3.8 (if one combines the 15 exposures together) and 10 (if one combines all the exposures of all objects together), respectively.

For each exposure of all objects, we estimated in each spectral window the residual local error after sky continuum subtraction (i.e., the accuracy of sky subtraction) as the relative mean between the object and sky spectra (divided by the mean sky). As we argued, considering all such measurements for all exposures of all objects in the sample simulates a 105-hour integration, providing us with well-defined statistics for measuring the mean expected accuracy as well as its associated uncertainty, as shown in Figure 3. We found in this case a mean accuracy of $0.6 \pm 0.2\,\%$. In comparison, the accuracy is degraded by a factor of ~ ten when using the simple staring-mode observations. This confirms the preliminary analysis conducted by Rodrigues et al. (2012), i.e., that cross beam-switching observations allow us to reach sky subtraction accuracies of a few tenths of a percent.

### Reaching the noise floor

We also investigated the exposure time needed to reach the noise floor at which the signal-to-noise ratio of the observations becomes limited by systematic effects associated to sky continuum subtraction inaccuracies. For this, we repeated the above measurements in samples of different sizes, which simulate different integration times, as argued above. Results are shown in Figure 4. At short integration times, the local residual error is dominated by random errors associated with the photon noise. It is expected from simple Poisson statistics that this error decreases as the square root of the integration time, until it reaches a floor (see Sharp & Parkinson, 2010). Given the limited size of our sample, it is difficult to measure such a decrease precisely, but a gradual decrease followed by a floor can indeed be seen in Figure 4. This floor is reached after 10–25 hours of integration at a value of 0.6%. At such large integration times, the local residual error starts to be dominated by systematic effects from the sky continuum subtraction.

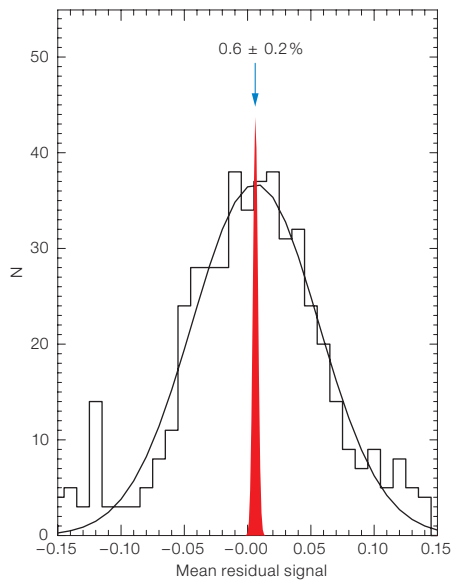

Figure 3. Histogram of the mean residual signal measured in each spectral window of all exposures of all objects observed with FLAMES–GIRAFFE. The mean value estimates the mean residual error signal, i.e., the accuracy of the sky continuum subtraction, obtained with simulated 105-hour long exposures. The red region represents the uncertainty associated with this value, which was derived using the standard error of the mean, i.e., as the standard deviation of the distribution divided by the square root of the sample size.

The trend shown in Figure 4 is similar to that found by Sharp & Parkinson (2010) at 600 nm, with a ~ 0.3% floor after 70–100 hours of integration. Besides, it is interesting to note that the 0.6% floor in residual local error is very close to the measured variations of the sky continuum background obtained by Puech et al. (2012) and Yang et al. (2012), which range between ~ 0.3 and 0.7%. This could indicate that long-exposure observations can really remove most of the instrumental inaccuracies and reach the physical limit due to sky continuum variations. The 0.3% floor found by Sharp & Parkinson (2010) could be due to the smaller variations in the 600 nm sky continuum, compared to the wavelengths around 900 nm that we are probing here.

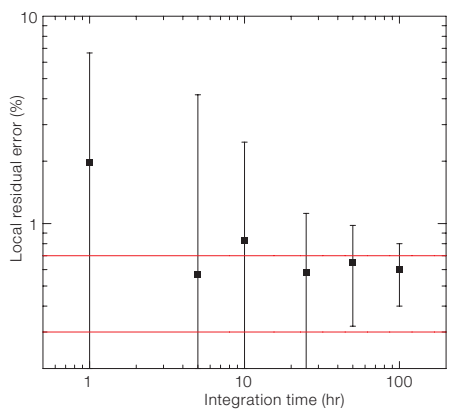

Figure 4. Local residual error as a function of integration time. The black line illustrates a decrease with the square root of integration time. The two horizontal red lines represent the range of variations found at ~ 10 arcsecond scales in the sky continuum background from Puech et al. (2012) and Yang et al. (2012).

It is important to recall that the results reported here were obtained on 1D spectra with non-optimal conditions. It is likely that using more advanced procedures in the data reduction (see Sharp & Parkinson, 2010) and with more control on the instrument design regarding the potential sources of inaccuracies detailed above, one might be able to lower the floor at which signal-to-noise ratio is limited by such systematic effects, and possibly to shorter integration times. Moreover, these results support the idea that controlling and measuring instrumental scattered light would remain the main obstacle to accurate spectroscopy of faint sources. We argued above that these results should apply up to *J*-band, but it will be important to confirm these results and characterise the sky continuum variations at even longer wavelengths, where the impact of scattered light keeps increasing.

The results reported here strongly suggest that the issue of the sky continuum subtraction is not a show-stopper for the study of very faint sources with fibre-fed spectrographs. Given the flexibility of these systems, it is likely that they will play a very important scientific role in the future generation of multi-object instruments such as MOONS for the VLT (Cirasuolo et al., 2011) or MOSAIC for the E-ELT (Evans et al., in prep.).